\documentclass[
    ,final            
  ]
  {aipproc}

\layoutstyle{6x9}

\begin{document}

\title{Dissipative processes in superfluid quark matter}


\author{Massimo~Mannarelli}{address={Departament d'Estructura i Constituents de
la Mat\`eria and Institut de Ci\`encies del Cosmos (ICCUB), Universitat de
Barcelona, Mart\'i i Franqu\`es 1, 08028 Barcelona, Spain}}

\author{Giuseppe~Colucci}{
  address={Universit\`a di
Bari, I-70126 Bari, Italia and I.N.F.N., Sezione di Bari,
I-70126 Bari, Italia }
}

\author{Cristina~Manuel}{
  address={Instituto de Ciencias del Espacio (IEEC/CSIC),
Campus Universitat Aut\`onoma de Barcelona, Facultat de Ci\`encies, Torre C5
E-08193 Bellaterra (Barcelona), Spain}}

\begin{abstract}
We present some results about dissipative processes in fermionic superfluids that are
relevant for compact stars. At sufficiently low
temperatures the  transport properties of a superfluid are dominated by phonons. We report the values of the bulk
viscosity, shear viscosity and thermal conductivity of phonons in quark matter at extremely high density and low temperature. 
Then, we present a new dissipative mechanism that can operate in compact stars and that is named ``rocket term". The effect of this dissipative mechanism on superfluid r-mode oscillations is sketched. 
\end{abstract}

\maketitle


\section{Introduction}
 Superfluidity is a property of quantum fluids related
with  the existence of low energy excitations  that satisfy the Landau's
criterion for superfluidity~\cite{Landau1941,IntroSupe,landaustat}
 \begin{equation}\label{landau-criterion}
{\rm Min \frac{\epsilon(p)}{p} \neq 0}\,,
\end{equation}
where $\epsilon(p)$ is the dispersion law of the excitations. The main property
of superfluids is that they have an extremely small viscosity and can
flow with very small friction. 

In general, superfluidity is due to the appearance of a  condensate which
spontaneously breaks a global  symmetry of the system. As a consequence, in the
low energy spectrum one has a Nambu-Goldstone boson,  $\varphi$, 
whose dispersion law  satisfies Eq.~(\ref{landau-criterion}). 

Superfluidity was first discovered in $^4$He, which becomes superfluid
when cooled at temperatures below $2.17$ K. The
superfluid property of  $^4$He atoms is due to the fact that they are bosonic with very
small interactions.  Therefore at low enough temperature they Bose-Einstein condense  in the lowest quantum state. 

Fermionic systems can become superfluid as well.  According with the Cooper
theorem, fermionic superfluidity takes place in quantum degenerate systems 
when the interaction between neutral fermions is attractive and the temperature
is sufficiently low.  In this case one has the formation of di-fermion
condensates, or   Cooper pairs, that behave
as  bosonic particle and  can condense. Also in this case superfluidity is
accompanied with the spontaneous breaking of a global
$U(1)$ symmetry and with the appearance of a  Goldstone boson.

Fermionic superfluids
might be present  in the interior of neutron stars where the temperature is low
and the typical energy scale of  particles is extremely high.   In particular,
in the inner crust of neutron stars the attractive interaction between neutrons
can lead to the formation of a BCS condensate.
Moreover, if deconfined quark matter is present in the core of neutron stars
it is very likely to be in a color superconducting phase~\cite{reviews}.

A means  to detect or to discard the presence of a superfluid
phase in compact stars
consists in studying the evolution of the r-mode oscillations~\cite{Andersson:2000mf}. R-modes are non-radial
oscillations of the star
with the Coriolis force acting as the restoring force. They provide a severe
limitation on the rotation frequency of the star through coupling to 
gravitational radiation. When dissipative phenomena damp these
r-modes the star can rotate without losing  angular momentum to gravitational radiation. If
dissipative phenomena are not strong enough,  these  oscillations  
will grow exponentially and the star will keep slowing down until some
dissipation mechanism  is able to  damp the r-mode oscillations. In this way one can put some
  constrains on  the stellar
structure ruling out  phases that do not have large enough viscosity.
For such studies it is necessary to consider in detail all the dissipative
processes.

\section{Dissipative processes in superfluids}
 The transport properties of a superfluid depend on the shear viscosity coefficient,  $\eta$, on 
\underline{three} independent bulk viscosity coefficients, $\zeta_1, \zeta_2,
\zeta_3$, and on the thermal conductivity $\kappa$. Microscopically, these 
coefficients are related with  interactions that take place in
the superfluid.  Macroscopically, they can be viewed
as the phenomenological coefficients which relate the rate of change of 
currents with the corresponding affinities~\cite{prigogine}. The requirement
that the dissipative processes lead to positive entropy production imposes
some constraint on these coefficients:  
$\kappa, \eta, \zeta_2, \zeta_3$   must be  positive and   $\zeta_ 1^2 \leq
\zeta_2 \zeta_3$.

The presence of three different  bulk viscosity  coefficients might sound
puzzling. It is related with the fact that the hydrodynamic equations governing a superfluid are essentially different from standard fluid equations. In a superfluid there  are two  independent motions, one normal and the other superfluid.
While the coefficient $\zeta_2$ plays the role of the standard bulk
viscosity coefficient, the coefficients  $\zeta_1$ and $\zeta_3$  provide a
coupling between the normal and superfluid components.
The friction forces due to bulk viscosities can be understood as drops, with
respect to their equilibrium values, in the main driving forces acting on the
normal and 
superfluid components. These forces are given by the gradients of the pressure
$P$ and of the chemical potential $\mu$. One can write in the comoving frame
\begin{eqnarray}
P &=& P_{\rm eq} - \zeta_1 {\rm div}(V^2 {\bf w}) -\zeta_2  {\rm div} {\bf u}\,,
\\
\mu &=& \mu_{\rm eq} - \zeta_3 {\rm div}(V^2 {\bf w}) -\zeta_1  {\rm div}{\bf
u}\, ,
\end{eqnarray}
where $P_{\rm eq}$ and $\mu_{\rm eq}$ are the equilibrium pressure and chemical
potential, $V$ is a quantity proportional
to the quantum condensate, $\omega^\mu = - \left(\partial^\mu \varphi + \mu
u^\mu \right)$ and $u^\mu$ is the  velocity of the fluid.

In a conformally invariant system it has been shown in Ref.~\cite{Son:2005tj}
that  $\zeta_1 = \zeta_2 =0$. However, $\zeta_3$, $\kappa$ and $\eta$  cannot be determined by the same symmetry argument. 

In the low temperature regime, $T \ll T_c$, where $T_c$ is the critical
temperature for superfluidity, the transport properties of superfluids are
determined  by  phonons. The contribution of   other degrees of freedom is
thermally suppressed.   In this case one can show that $\zeta_1^2= \zeta_2
\zeta_3$, meaning that there are only two independent bulk viscosity
coefficients.

For $T \ll T_c$ the transport coefficients strongly depend on the phonon
dispersion law. 
The shear viscosity is the only transport coefficient that does not vanish for
phonons with a linear dispersion law. But, for the bulk viscosities and for the
thermal conductivity  one  has to include the term cubic in momentum 
\begin{equation}\label{dispersion}
\epsilon(p)= c_s p + B p^3  + {\cal O}(p^5)\,.
\end{equation}
Moreover, in the computation of the bulk viscosity one has to consider the
processes that change the number of phonons. The parameter $B$  determines
whether  some processes are or are not kinematically allowed.  For $B>0$ the
leading contribution comes from the Beliaev process  $\varphi \to \varphi \varphi$.
In the opposite  case the Beliaev process is not kinematically allowed and one
has to consider the processes  
$\varphi \varphi \to \varphi \varphi \varphi$. 

A different dissipative process is associated with the
change in the number densities of the various species present in the superfluid,
which leads to the appearance of the so-called rocket term in the Euler equations of the system.
This force is  due to the fact that when two or more fluids  move with different
velocities a  change of one component into the other results in a 
 momentum transfer between the fluids. This change in momentum is not
reversible, because it is always the faster moving fluid that will lose
momentum.  The resulting dissipative force is  proportional to the mass rate
change, and to the relative velocity of the fluids.
The name ``rocket term" reminds that the same phenomenon  takes place  
in the dynamical evolution of a rocket whose mass is changing in time as it
consumes its fuel. In compact stars number changing  processes can take place  in the
outer core and in the inner crust of the star and are related
to the Urca processes and/or interactions between the superfluid and the crust. 
The dissipative force due to the  rocket
term has  been considered in the context of r-mode
oscillations of standard $npe$ neutron stars, for the first time  in Ref.~\cite{Colucci:2010wy}.

\section{Cold and dense quark matter}

Quantum Chromodynamics predicts that
at asymptotically high densities  quark matter is in the color-flavor locked (CFL)
phase~\cite{Alford:1998mk}.
In this phase up, down and strange quarks of all three colors pair forming a
di-fermion condensate that  breaks the  $U(1)_B$ symmetry spontaneously.

At asymptotic densities, the dispersion law of phonons in the CFL phase  has been derived in~\cite{Zarembo:2000pj},  and one has that $
c_s= \sqrt{\frac 1 3}$  and $ B= -\frac{11 c_s}{540 \Delta^2}$, 
where  $\Delta$ is the gap parameter. The effect of the quark masses can be taken into account  perturbatively,  as far as $m_s \ll \mu$,
moreover the coupling constant is small, $g (\mu) \ll 1$, and one can assume
that CFL is approximately scale invariant. 

A  first study of the shear viscosity and of the contribution to the bulk
viscosity coefficients  due to phonons has been presented in
Refs.~\cite{Manuel:2004iv,Manuel:2007pz,Mannarelli:2009ia}. Beside phonons, kaons  may give a
sizable contribution to the transport coefficients. The contribution of kaons to
$\zeta_2$  has  been studied in Ref.~\cite{Alford:2007rw}.  There is still no
computation of the contribution of kaons to the remaining bulk viscosity
coefficients. 

Considering only the contribution of phonons, in Ref.~\cite{Mannarelli:2009ia}
it has been found that neglecting the running of the QCD gauge coupling
\begin{equation}
\zeta_1 \sim \frac{m_s^2 \mu^7}{T \Delta^8}\,, \qquad  \zeta_2\sim \frac{m_s^4 \mu^8}{T \Delta^8}\,,  \qquad \zeta_3 \sim
\frac{1}{T} \frac{\mu^6}{\Delta^8} \, .
\end{equation}
Notice that  for vanishing quark masses the system is scale invariant and $\zeta_1$ and $\zeta_2$
vanish.

The contribution to thermal conductivity due to phonons and kaons has recently
been studied in Ref.~\cite{Braby:2009dw}.  The thermal conductivity from phonons
turns out to be dominant and given by 
\begin{equation}
\kappa \sim 6 \times 10^{-2} \frac{\mu^8}{\Delta^6}\,,
\end{equation}
whereas the shear viscosity from phonons as determined in~\cite{Manuel:2004iv}
is given by 
\begin{equation}
\eta \simeq 1.3 \times 10^{-2} \frac{\mu^8}{T^5} \,.
\end{equation}

A different dissipative mechanism is due to mutual friction, that has been studied in the CFL phase
in Refs.\cite{Mannarelli:2008je} and \cite{Andersson:2010sh}.

\section{Rocket term}

For a system consisting of neutrons,  protons  and  electrons,
the mass conservation law is given by, see {\it
e.g.}~\cite{Prix:2002jn},
\begin{equation}\label{continuity}
 \partial_t\rho_x+\nabla_i(\rho_x v_x^i)= \Gamma_x \ ,
\end{equation}
where $\Gamma_x$ is the particle mass creation rate per unit volume and the 
index $x=n,p,e$ refer to the particle species, that is, neutrons, protons and
electrons. In these equations we have  considered that some process  can
convert neutrons in protons and electrons and {\it vice versa}.
Therefore, we are assuming that the various components are not separately
conserved. One possible mechanism leading to a change in 
the particle number densities is  given by the Urca process 
\begin{equation}
n  \rightarrow  p + e^- + {\bar \nu}_e \, , 
\qquad
p + e^- \rightarrow n + \nu_e \,.
\end{equation}

A different process is the {\it crust-core transfusion},
a process that takes place when  the ionic
constituent of the crust are squeezed by the underlying superfluid and part of their  content is
released and augments the superfluid  component.  The opposite mechanism, related to
a reduction of the pressure leads to the neutron capture by the ions of the
crust. We restrict our analysis to the beta decay
processes. 

The three particle creation rates are not independent quantities,
because charge and baryon number conservations imply that
$\Gamma_e = \Gamma_p  -\Gamma_n$, meaning that only one creation rate is independent.

It is possible to simplify the treatment of the system considering that 
electrons and protons are locked together by the electromagnetic interaction. Thus 
they move with the same velocity. Moreover, charge neutrality implies
that $n_e = n_p$.
Therefore, electrons and protons can be treated as a single charge-neutral fluid
and henceforth we shall refer to this fluid as the ``charged'' component.
The two fluids have mass densities 
\begin{equation}
\rho_n = m_n\, n_n \qquad {\rm and } \qquad \rho_c = m_n\, n_c \,, 
\end{equation}
where $m_n=m_p + m_e$ and $n_e=n_p=n_c$. The Euler equations obeyed by the two
fluids are given by 
\begin{eqnarray}
\label{Eul-neutron}
 (\partial_t +v_n^j\nabla_j)(v_i^n+\epsilon_n
w_i)+\nabla_i(\tilde{\mu}_n+\Phi)+\epsilon_n w^j\nabla_iv_j^n
& = & 0  \,,\\
\label{Eul-proton}
(\partial_t +v_c^j\nabla_j)(v_i^c - \epsilon_c w_i)+\nabla_i(\tilde{\mu}_c+\Phi)
-\epsilon_c w^j\nabla_iv_j^c & = & \left(1
-\epsilon_n -\epsilon_c  \right) \frac{\Gamma_n}{\rho_c} w_i \,,
\end{eqnarray}
where $\Phi$ is the gravitational potential,  $i,j$ label the space components;  we have defined a chemical potential
by mass $\tilde{\mu}_x = \mu_x/m_n$,  and 
 ${\bf w} = {\bf v}_c -{\bf v}_n$ represents the relative velocity between the
two fluids.  The quantities   $\epsilon_n$ and $ \epsilon_c$ are 
the entrainment parameters, that are related to the fact that momenta and
velocities of quasiparticles may not be aligned~\cite{Andreev}.

The term on the right hand
side of Eq.~(\ref{Eul-proton}) is the rocket term.  
In the analysis of the possible dissipative mechanisms of star oscillations
this term is usually neglected.  Indeed, it is in general assumed that the
neutron, proton and electron numbers are separately conserved quantities, that
is $\Gamma_p = \Gamma_e =\Gamma_n =0$. In Ref.~ \cite{Colucci:2010wy} we have considered the effect of this term and obtained the corresponding damping timescale, assuming that the change in the number densities is due to direct Urca processes.

\subsection{Stability window for superfluid r-modes }
\begin{figure}
\includegraphics[height=.22\textheight]{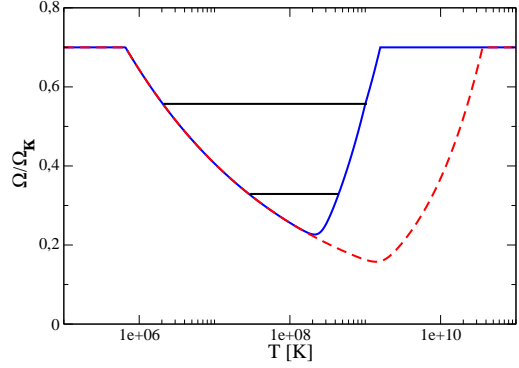}
\caption{Instability window of the superfluid r-modes of a star with
uniform density $\rho=2.5 \rho_0$, with radius $R=10$ km and 
mass $M\simeq 1.47 M_\odot$. The dashed red line represents the
instability window in the absence of the
rocket term. The full blue  line represents the instability window with the
inclusion of the rocket term.
The horizontal full
lines correspond to the effect of the mutual friction for $\epsilon=0.0002$,
lower line, and $\epsilon=0.002$ upper line. In our simplified model of star,
the mutual friction is independent of the temperature. 
}
\label{fig:instability}
\end{figure}
In Fig.~\ref{fig:instability} we report the result for the superfluid r-mode
``instability window" for a star with uniform density, $\rho = 2.5 \rho_0$, and
$R = 10$ km. It is obtained comparing  the various dissipative timescales with the growth timescale of gravitational wave radiation. Stars above the various lines are unstable.
The dashed red line represents the instability
window in the absence of the rocket term. The region above
the dashed red line is unstable  when
only shear and bulk viscosity damping  are considered.
At low temperature, shear viscosity is the dominant dissipative
mechanism. With increasing temperature shear viscosity is less efficient and
it can damp r-mode oscillations for smaller and smaller values of the
frequency. The behavior of bulk viscosity is the opposite and starts to damp
r-mode oscillations for temperatures of the order of $10^{10}$ K.
The full blue  line  represents the effect of the rocket term. It has a
behavior qualitatively similar to the bulk viscosity, but it becomes
effective at smaller temperatures. Therefore, the instability window with the
inclusion of the rocket term is much reduced. The full horizontal lines correspond to
the effect of mutual friction, see e.g.~\cite{Lee:2002fp}, for two different values of the entrainment parameter.

\begin{theacknowledgments}
This work has been supported in part by  the INFN-MICINN grant 
with reference number FPA2008-03918E. The work of CM has been supported  by the
Spanish  grant FPA2007-60275. The work of MM has been
supported by the Centro Nacional de F\'isica de Part\'iculas, Astropart\`iculas
y Nuclear (CPAN) and by the Ministerio de Educaci\'on y Ciencia (MEC) under
grant AYA 2005-08013-C03-02, FPA2007-66665 and 2009SGR502.
\end{theacknowledgments}



\bibliographystyle{aipproc}   





\end{document}